\documentclass[superscriptaddress,prl,showpacs,twocolumn]{revtex4-1}
\usepackage{psfrag}
\usepackage{amssymb,amsmath,amsthm,color}
\usepackage{graphicx}

\newcommand{\beq}{\begin{equation}}
\newcommand{\eeq}{\end{equation}}

\begin{document}
\title{Self-assembly of catalytically active colloidal
molecules: Tailoring activity through surface chemistry}

\author{Rodrigo Soto}
\affiliation{Departamento de F\'{\i}sica, Facultad de Ciencias F\'{\i}sicas y Matem\'aticas Universidad de Chile,
Av. Blanco Encalada 2008, Santiago, Chile}
\affiliation{Rudolf Peierls Centre for Theoretical Physics, University of Oxford, Oxford OX1 3NP, UK}

\author{Ramin Golestanian}
\affiliation{Rudolf Peierls Centre for Theoretical Physics, University of Oxford, Oxford OX1 3NP, UK}

\date{\today}

\begin{abstract}

A heterogeneous and dilute suspension of catalytically active colloids is
studied as a non-equilibrium analogue of ionic systems, which has the
remarkable feature of action-reaction symmetry breaking. Symmetrically
coated colloids are found to join up to form self-assembled molecules that
could be inert or have spontaneous activity in the form of net translational
velocity and spin depending on their symmetry properties and their
constituents. The type of activity can be adjusted by changing the surface 
chemistry and ambient variables that control the surface reactions and 
the phoretic drift.

\end{abstract}


\maketitle

In a solution that maintains a gradient in the concentration
of a solute, colloidal particles could be driven up or down
the gradient. Since the pioneering work of Derjaguin and 
coworkers \cite{Derjaguin}, this nonequilibrium transport phenomenon, 
which is termed as diffusiophoresis, has been studied
in detail \cite{Anderson-review,JulProEPJE,Bocquet2,Bocquet3}.
The gradients that are needed to drive the system away from equilibrium
could be generated through catalytic activity, which can be designed
to take place asymmetrically on the surface of the colloids themselves, leading
to self-propulsion \cite{gla2005,ruckner2007,gla2007,popescu3,udo,lauga}
and anomalous stochastic dynamics \cite{MSD}. Recent experiments on self-phoretic 
active colloids have helped characterize their properties and highlight their 
potential as a versatile means of studying active nonequilibrium systems
\cite{Jon07,showalter,ozin,steve1,lyderic2,Bech,steve2,steve3,Bech2,pine,RayRev}.
Self-diffusiophoresis has been predominantly studied in
the context of artificial micro- and nanoscale swimmers,
which is currently a very active field of research
\cite{SynthMot1,SynthMot2,SynthMot3,Sen2012}.

The non-equilibrium behavior of catalytic colloids is characterized
by two independent parameters. First, the surface activity $\alpha$,
which describes the production or consumption  of chemicals at the colloidal
surfaces. For the symmetrically coated colloids we consider here, the concentration profile
$C$ is obtained by solving the diffusion coefficient $\nabla^2 C=0$ with the boundary
condition $- D \partial_r C|_{r=R}=\alpha$, where  $R$ is the radius of the colloid
and $D$ is the effective diffusivity of the chemicals. The activity results in a
concentration profile that decays as $C(r) \sim {\alpha R^2}/{(D r)}$.
The second parameter is the surface mobility $\mu$, which determines how
the colloid moves in response to the chemical gradients generated by the colloid
itself or by other active colloids in the medium \cite{gla2007,note}.
Both parameters can be positive or negative (or zero), independently, and can be
in principle controlled by modifying the surface chemistry of the colloids.

\begin{figure}[t!]
\includegraphics[width=\columnwidth]{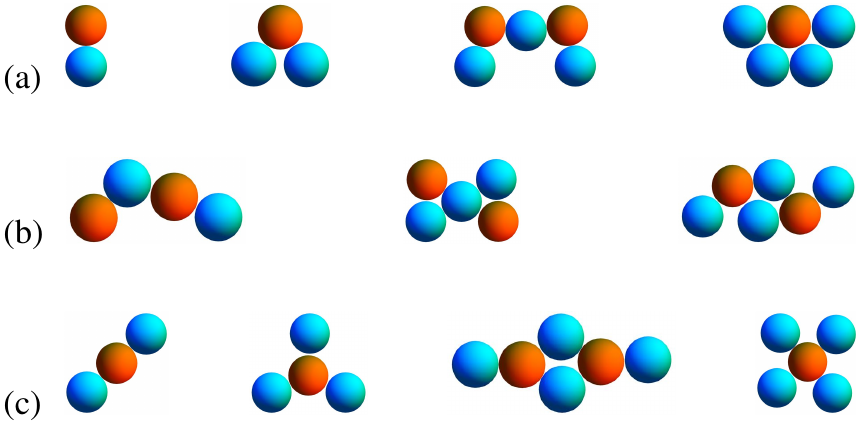}
\caption{(color online).
Examples of self-assembled active molecules.
(a) Self-propelled molecules moving up ($A$ in orange/dark gray and $B$ in blue/light gray). From left to right:
$AB$ ($\widetilde{\alpha}_A=2$, $\widetilde{\mu}_A=0$);
$AB_2$ ($\widetilde{\alpha}_A=2$, $\widetilde{\mu}_A=0$);
$A_2B_3$ ($\widetilde{\alpha}_A=2$, $\widetilde{\mu}_A=0$);
$AB_4$ ($\widetilde{\alpha}_A=5$, $\widetilde{\mu}_A=2$).
(b) Self-rotating molecules (spinning in the clockwise direction). From left to right:
$A_2B_2$ ($\widetilde{\alpha}_A=1$, $\widetilde{\mu}_A=3$);
$A_2B_3$ ($\widetilde{\alpha}_A=3$, $\widetilde{\mu}_A=1$);
$A_2B_4$ ($\widetilde{\alpha}_A=2$, $\widetilde{\mu}_A=1$).
The first one is polar and therefore exhibits spontaneous translation and rotation.
The two other cases are non-polar and therefore they rotate without directed motion.
(c) Inert molecules. From left to right:
$AB_2$ ($\widetilde{\alpha}_A=1$, $\widetilde{\mu}_A=3$);
$AB_3$ ($\widetilde{\alpha}_A=2$, $\widetilde{\mu}_A=1$);
$A_2B_4$ ($\widetilde{\alpha}_A=2$, $\widetilde{\mu}_A=3$);
$AB_4$ ($\widetilde{\alpha}_A=2$, $\widetilde{\mu}_A=0$).
}
\label{fig.stablemolecs}
\end{figure}

A single spherical colloid that is symmetrically coated does not self-propel
but phoretic interaction between two such colloidal particles could lead to net motion.
Within the far-field approximation, the drift velocity of particle 2 due to the activity
of particle 1 is given as
$\vec{V}_2 = -\mu_2 \nabla C|_{2} \sim \alpha_1 \mu_2 \frac{R^2 \vec{r}_{12}}{D |\vec{r}_{12}|^3}$,
where $\vec{r}_{12}=\vec{r}_2-\vec{r}_1$,
which is different from the drift velocity of particle 1,
$\vec{V}_1 = -\mu_1 \nabla C|_{1} \sim -\alpha_2 \mu_1 \frac{R^2 \vec{r}_{12}}{D |\vec{r}_{12}|^3}$.
The fact that the action-reaction symmetry is broken implies that the whole pair moves with
a velocity proportional to $(\alpha_1\mu_2-\alpha_2\mu_1)$, while the two colloids attract
or repel each other with a relative velocity proportional to $(\alpha_1\mu_2+\alpha_2\mu_1)$.
Due to this nonequilibrium property, it is possible to vary the parameters such that we
obtain stable compounds with various types of activity (see Fig. \ref{fig.stablemolecs}) 
via self-assembly. This is what we study in this Letter.

The effective interaction between active colloids is similar to unscreened electrostatic
interaction between charged colloidal particles in a fluid. However, we need to generalize
the concept of charge to differentiate between the charge that produces the field, $\alpha$,
and the charge that responds to the field, $\mu$. The presence of two types of charge
gives rise to non-intuitive effects absent in electrostatic systems. For example,
if $\alpha_1 > 0$ and $\mu_1 >0$ for particle 1, and $\alpha_2 > 0$ while $\mu_2 <0$
for particle 2, 1 will be repelled by 2 but 2 will be attracted by 1. We can categorize
the different possible scenarios as follows, if only the signs of the two charges
($\alpha$,$\mu$) are taken into account. In a mixture of $(++)$ and $(--)$ particles,
unequal particles attract and equal particles repel, similar to electrostatics.
In a mixture of $(+-)$ and $(-+)$ particles, equal particles attract and unequal
particles repel, which is opposite to the electrostatic rules. More complex situations
can be obtained by mixing different kind of particles. In such a mixture, we generically have 
$\alpha_A \mu_B \neq \alpha_B \mu_A$, and hence nonequilibrium colloidal activity. 
Only fine tuning could lead to equilibrium colloidal dynamics for dissimilar colloids 
(although chemical activity will always be present). Note that a mixture of particles 
of equal type but different radii also breaks action-reaction symmetry, but exhibits 
a more restricted phenomenology due to the signs of the charges being fixed. Here, 
we will concentrate on a $(++)$--$(--)$ mixture, and adopt the notation of $A=(++)$ 
and $B=(--)$. This choice allows the self-assembly of different types of particles, 
and thus, leads to molecules that exhibit activity.

We have performed Brownian dynamics simulation on mixtures of $A$ and $B$ colloids and studied
the self-assembly dynamics, and in particular, the spontaneous formation of active molecules. Besides
the phoretic interaction, we have incorporated steric forces by treating the colloids as hard spheres
all having the same diameter $\sigma$. The addition of the hard core repulsion makes this model similar
to the restricted primitive model (RPM) for charged colloids \cite{YanLevin}, and allows the formation
of stable compact colloidal molecules. We have numerically solved the following stochastic equations
\begin{equation}
\frac{d \vec{r}_i}{d t}=\sum_{k\neq i} \frac{\alpha_k \mu_i R^2}{6 \pi D} \frac{\vec{r}_{ki}}
{|\vec{r}_{ki}|^3} +\vec{\xi}_i(t)=V_0  \sum_k \widetilde{\alpha}_k\widetilde{\mu}_i
\frac{\sigma^2 \vec{r}_{ki}}{|\vec{r}_{ki}|^3} +\vec{\xi}_i(t). \label{eq.fullmodel}
\end{equation}
The first term corresponds to the pair phoretic interactions in the far-field approximation
and $\vec{\xi}$ is a random velocity representing white noise of intensity $2 D_c$ with $D_c$ being
the (passive) diffusion coefficient of the colloidal particles, which is proportional to the temperature.
Using $\alpha_0$ and $\mu_0$ as characteristic values of the surface parameters, we have made them
dimensionless by defining $\widetilde{\alpha}=\alpha/\alpha_0$ and $\widetilde{\mu}=\mu/\mu_0$. This gives
us the overall velocity scale of $V_0=\frac{\alpha_0 \mu_0 R^2}{6 \pi D\sigma^2}=\frac{\alpha_0 \mu_0}{24 \pi D}$.

To simulate the model we adopt a similar strategy as the one described in \cite{Brownian}.
The system is advanced in time steps $\delta t=0.001\sigma/V_0$. At each time step
the drift velocity for each particle is computed summing all pairwise interactions.
Then, particles are advanced using a forward Euler scheme that adds the Brownian
displacement associated to the time step. At this stage pairs of overlapping particles
are reflected, in order, by the same distance they were overlapped. The procedure is repeated until
there are no remaining overlaps. This procedure is known to reproduce, in the limit
of small time steps, the dynamics of hard core systems \cite{Brownian}. Periodic boundary conditions
are used and interactions are treated using the minimal image convention \cite{Allen,FS}.
Although interactions are long-ranged, for the purpose of this study that focuses on
the formation of short scale structures, the Ewald summation scheme was not required.
For simplicity, simulations are in 2D, but the interaction is computed considering that solute molecules
diffuse in three dimensions, with the effective potential decaying as $1/r$. This potential,
instead of the strictly 2D $\log r$ form has the advantage that the electrolyte phenomenology
is the same as in 3D and there is no global Kosterlitz-Thouless transition.

We define the dimensionless area fraction $\nu=N\pi \sigma^2/4L^2$, where $N$ is
the total number of particles and $L$ is the system size, and the dimensionless noise intensity
$\widetilde D_c=D_c/V_0\sigma$, which we regard as a dimensionless temperature.
$\alpha_0$ and $\mu_0$ are chosen such that for $B$ particles $\widetilde{\alpha}_B=\widetilde{\mu}_B=-1$.
Without loss of generality, we choose $\widetilde{\alpha}_A\geq 1$ (the opposite case is obtained by
exchanging the role of $A$ and $B$ particles). Since our interest is in the formation of active aggregates
composed of a few particles, simulations are performed in the dilute regime ($\nu\ll 1$).
Simulations are performed with $N=1200$ and $L/\sigma=1200-3000$, leading to area fractions in the range
$\nu=(1.0 - 6.5)\times10^{-4}$.
A stationary state (equilibrium or nonequilibrium) can only be reached in the system if there is no net production
or consumption of chemicals. This condition requires $N_A\alpha_a+N_B\alpha_B=0$, which is equivalent to global
neutrality in our electrostatic analogy. Consequently, the relative number of particles is fixed by the relation
$N_B/N_A=\widetilde{\alpha}_A$. The initial conditions are such that both species are mixed homogeneously at random.
Finally, two particles belong to the same molecule if their relative distance is smaller than $1.5\sigma$.

At low area fraction and high temperatures, we observe that stable low weight molecules
are formed. Some of these molecules are shown in Fig. \ref{fig.stablemolecs}. Depending on
the symmetry of the self-assembled molecules, they could exhibit different types of activity,
in the form of overall translational and/or rotational self-propulsion. Increasing the area
fraction or reducing the temperature leads to the formation of large molecules or clusters.
The simplest molecule is $AB$. It has a spontaneous translational velocity in all
non-equilibrium conditions. The next molecule that appears is $AB_2$, which could
acquire two different geometric configurations or {\it isomers}: (i) if the molecule
is linear, by symmetry, it does not self-propel; (ii) if the molecule adopts a triangular
shape it swims steadily. The simplest self-rotating molecule that does not
show simultaneous self-propulsion is $A_2 B_3$. Other stable molecules are formed,
with varying types of activity (see Fig. \ref{fig.stablemolecs}).

\begin{figure}
\includegraphics[width=.9\columnwidth]{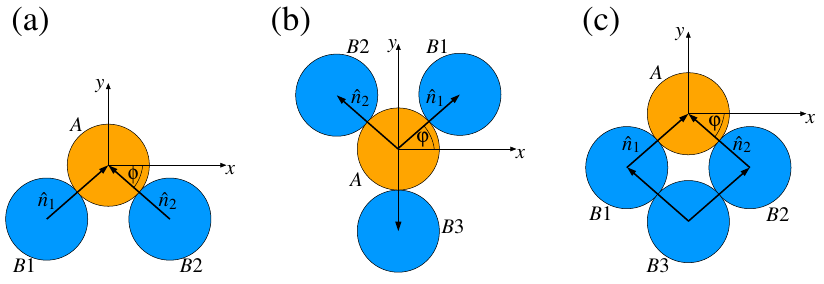}
\caption{(color online). Geometric parametrization of the $AB_2$ and $AB_3$ molecules used in their
stability analysis. (a) The angle $\phi$ parameterizes the deviation from a linear
chain of the $AB_2$ molecule. (b) The angle $\varphi$ parameterizes the deformation
of the $AB_3$ molecule from the symmetric star configuration. (c) The angle
$\varphi$ parameterizes the detachment of one $B$ colloid from the $AB_3$ molecule.}
\label{fig.geomAB2}
\end{figure}

The statistical properties of the $AB$ molecule are obtained by considering
the velocities of each colloidal particle [Eq. (\ref{eq.fullmodel})].
Changing variables to the relative distance $\vec{\rho}=\rho \hat{\rho}$
and the center of mass position $\vec{R}$ (and omitting the noise terms),
we obtain for the velocities of the center of mass $\vec{V}_R$ an the relative distance $\vec{V}_\rho$
\begin{eqnarray}
&& \vec{V}_R=V_{0} (\widetilde{\mu}_A-\widetilde{\alpha}_A) \frac{\sigma^2}{2 \rho^2}
\hat{\rho}, \label{eq.VCMAB1}\\
&& \vec{V}_\rho=2 U(\rho) \hat{\rho} + V_0 \sigma^2 (\widetilde{\mu}_A+\widetilde{\alpha}_A)
\frac{\hat{\rho}}{\rho^2},\label{eq.VCMAB2}
\end{eqnarray}
where $U$ models the hard core repulsion. The propulsion speed of the $AB$ molecule 
increases as $\rho$ is decreased, achieving the maximum value when the particles are
in contact. The corresponding Fokker-Planck equation [when considering the noise terms
in Eqs. (\ref{eq.VCMAB1}) and (\ref{eq.VCMAB2})] can be simplified by separation
of variables since the equation for $\rho$ is completely decoupled, and can be solved
at the stationary state to yield the 2D pair distribution function
\begin{equation}
P(\rho)=\left\{\begin{array}{lcl} A \rho e^{\ell/\rho} &;& \rho>\sigma \\
0 &;& \rho\leq \sigma
 \end{array}\right.
\end{equation}
where $A$ is a normalization constant. Here, $\ell=V_0 \sigma^2(\widetilde\mu_A  + \widetilde \alpha_A)/D_c$
is the analogue of the Bjerrum length, which corresponds to the distance at which the distribution
function is minimum and marks an energy-entropy crossover. In ionic liquids, this length is used
to divide between bound states and free (or ionized) states \cite{YanLevin}.
We can use $P(\rho)$ to obtain the distribution for the center of mass (propulsion) velocity
using Eq. (\ref{eq.VCMAB1}), which yields
\begin{equation}
f(V_R)=\frac{f_0}{V_R^2} \exp\left\{\frac{\ell}{\sigma} \sqrt{\frac{2 V_R}{V_{0}
|\widetilde{\mu}_A-\widetilde{\alpha}_A|}} \right\},\label{eq.fV}
\end{equation}
where $f_0$ is a normalization constant. The distribution is valid for velocities up to
the upper cutoff (maximum value) of $V_{0} |\widetilde{\mu}_A-\widetilde{\alpha}_A|/2$.
Our simulations in the studied density range give results in agreement with Eq. (\ref{eq.fV}).

\begin{figure}[t!]
\includegraphics[width=0.99 \columnwidth]{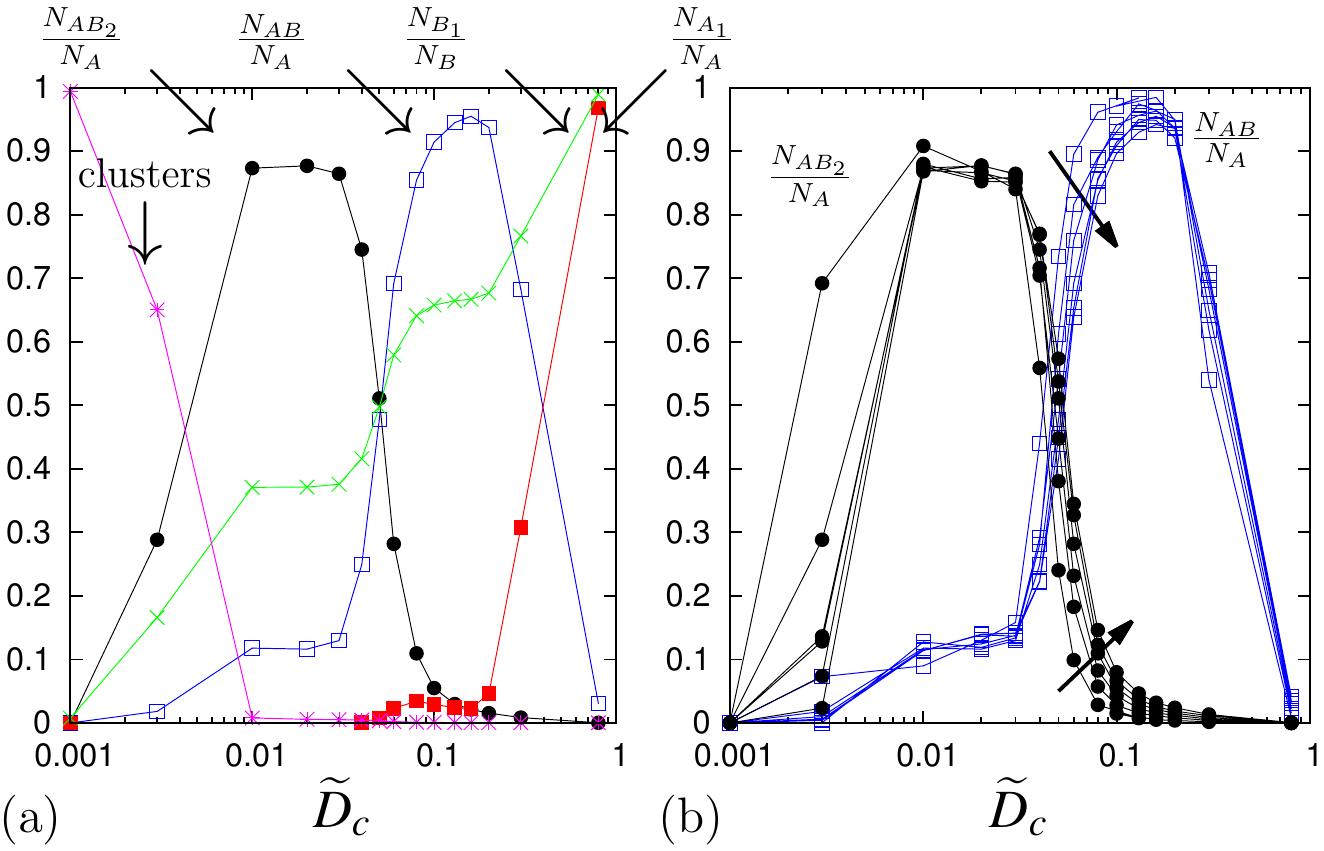}
\caption{(color online). Relative populations of different types of molecules as a function of the dimensionless diffusion coefficient
(that is proportional to the temperature) in the stationary state. (a) Relative populations for all observed structures
for area fraction $\nu=4.36\times10^{-4}$. For the {\em clusters}, the relative total mass defined as
$\sum_{m,n}^* (m+n) N_{A_n B_m}/[N_A+N_B]$ is plotted (rather than relative population), where $*$ means that in the sum
the following cases are excluded $(n,m)=(1,0), (0,1), (1,1), (1,2)$. (b) The populations of $AB_2$ and $AB$ molecules
for different area fractions varied in the range $\nu= (1.0 - 6.5)\times10^{-4}$, increasing in the direction shown by the arrows.
The thin lines are drawn to guide the eye. In the simulations $\widetilde{\alpha}_A=3$ and $\widetilde{\mu}_A=0$.}
\label{fig.concentrationABn}
\end{figure}

The equilibrium configuration of the $AB_2$ molecule, which can exist in two stable
states or isomers as mentioned above, can be obtained using a kinematic analysis.
Consider a molecule in which a central $A$ colloid is in contact with two $B$ colloids
with a bond angle $\phi$ as shown in Fig. \ref{fig.geomAB2}. The deterministic velocities
of the three particles [Eq. (\ref{eq.fullmodel})] are given by the following expressions
\begin{eqnarray}
&&\vec{V}_A = -V_0 \widetilde{\mu}_A \hat{n}_1
- V_0 \widetilde{\mu}_A \hat{n}_2 +U \hat{n}_1 + U\hat{n}_2, \label{AB2-1} \\
&&\vec{V}_{B1}= V_0 \widetilde{\alpha}_A \hat{n}_1 -U\hat{n}_1
- \frac{V_0}{(2\cos\phi)^2} \hat{x},\label{AB2-2} \\
&&\vec{V}_{B2}=V_0 \widetilde{\alpha}_A \hat{n}_2 -U\hat{n}_2
+ \frac{V_0}{(2\cos\phi)^2} \hat{x},\label{AB2-3}
\end{eqnarray}
using the same notation to model the hard core repulsion. The unit vectors ($\hat{n}_1$,
$\hat{n}_2$, and $\hat{x}$) are defined in Fig. \ref{fig.geomAB2}. Imposing the constraint
that the  relative $AB$ distances be kept fixed ($\vec{V}_A-\vec{V}_{Bi})\cdot\hat{n}_i=0$,
the deflection angle satisfies the equation
\begin{eqnarray}
\dot{\phi}=\frac{2V_0\sin\phi}{\sigma(2-\cos 2\phi)}\left [(\widetilde{\alpha}_A-\widetilde{\mu}_A) \cos\phi
-\frac{3}{8\cos^2\phi}\right].  \label{eq.phi-1}
\end{eqnarray}
The first term in the above equation is due to the activity, whereas the second term is due
to the repulsive interaction among the $B$ particles, which favors a linear chain ($\phi=0$).
When $(\widetilde{\alpha}_A-\widetilde{\mu}_A)< 3/8$, $\phi=0$ is stable and the molecule
adopts the non-propelling linear isomer configuration. In the opposite case, when
$(\widetilde{\alpha}_A-\widetilde{\mu}_A)> 3/8$, any small initial angle is destabilized
and the final angle depends on the value of the parameters. In this case the molecule adopts
the triangular isomer configuration and swims in the positive $\hat{y}$ direction with a velocity
that increases with $\phi$. Replacing the equilibrium angle in Eqs. (\ref{AB2-1})-(\ref{AB2-3}),
yields the swimming velocity 
$V_{AB_2}=V_0\sqrt{4(\widetilde{\alpha}_A-\widetilde{\mu}_A)^2-3^{2/3}(\widetilde{\alpha}_A-\widetilde{\mu}_A)^{4/3}}/3$.
One could induce a transition between the isomers and  modify the molecule activity by changing the values of the charges, which depend on the temperature and other ambient variables.

A similar analysis for $AB_3$ molecule, when $\widetilde{\alpha}_A=3$ and $\widetilde{\mu}_A=0$,
reveals that it is unstable, and that it will break up into an $AB_2$ molecule and a free $B$
colloid, through a detachment process characterized by the steady increase of the angle $\varphi$
as shown in Figs. \ref{fig.geomAB2}b and \ref{fig.geomAB2}c. In the case of $\widetilde{\alpha}_A=4$
and $\widetilde{\mu}_A=2$, it is possible to have stable self-propelling $AB_3$ molecules,
although they fluctuate strongly and the swimming direction changes rapidly. A movie of both
possibilities is shown in the Supplemental Material \cite{supmat}.

It is interesting to note that despite the condition of global neutrality and in contrast
with what we expect from ionic bonds, small stable molecules are not necessarily neutral.
Moreover, neutral molecules could even be unstable as it is the case for $AB_3$.
One can argue that although non-neutral molecules attract monomers and other molecules,
when they are active they remain charged because their propulsion strongly hinders
the capture process and by the detachment of colloids, as can be observed in the Supplemental 
Material movies~\cite{supmat}.

The stationary distribution of molecules is studied for $\widetilde{\alpha}_A=3$
and $\widetilde{\mu}_A=0$, in which case the $AB$ an $AB_2$ molecules are active.
The number of these molecules is presented in Fig.~\ref{fig.concentrationABn}
for simulations done with $N=1200$ and various densities and temperatures. We 
observe that in the steady state the system forms small molecules with the overwhelming 
majority being of the form $AB_n$ ($>98\%$ for $n\leq4$) and no large clusters, except 
for very low temperatures ($\widetilde D_c\lesssim 0.01$) as shown in Fig. \ref{fig.concentrationABn}a. The types of 
self-assembled molecules are predominantly controlled by temperature, with little 
dependence on density in the range studied as Fig. \ref{fig.concentrationABn}b shows. 
Between $\widetilde D_c=0.015$ and $\widetilde D_c=0.05$ there is a sharp transition 
from $AB_2$ to $AB$. This observation is generic and not specific to this particular 
case. For example, in the case of $\widetilde{\mu}_A=1$, a similar behavior is
obtained but the transition temperature is located between $\widetilde D_c=0.05$
and $\widetilde D_c=0.15$. 

Note that although the relative concentration of molecules 
as a function of temperature follows a similar trend as in equilibrium, contrary to 
the equilibrium case the stationary distributions cannot be computed in terms of 
effective molecular free energies or equilibrium constants. If this were the case, the relative concentrations 
would satisfy $N_{AB_2}=K N_{B_1} N_{AB}$ at low densities, with $K$ the equilibrium constant being a function 
of the temperature only. This and similar relations for the other molecules, however, are not quantitatively satisfied here.

We have studied a mixture of active colloids with asymmetric mutual diffusiophoretic interactions
and shown that they self-assemble into active molecules. Moreover, we have found that
the self-assembled structures are stable (distinct like molecules that form from ions of given
valence), exhibit novel non-equilibrium dynamics, and present different types
of activity that can be varied by changing simple parameters. To capture
the main features of the self-assembled molecules we simplified the colloidal interaction
by only considering the far-field contribution (effective potential decaying as $1/r$).
To make quantitative predictions, corrections at short distances should be included
to consider the exact solution of the diffusion equation with the appropriate boundary
conditions, and the depletion of chemicals near the colloidal surfaces. However, 
we expect that these corrections will only modify the result quantitatively, and 
that the main qualitative features of the above phenomenology would remain the same. 
Hydrodynamic interactions between colloids have not been considered. Since they respect 
the action-reaction symmetry, we do not expect them to introduce qualitative changes 
in our description except for the suppression of relative slip between colloids due 
to the lubrication layer and possible quantitative changes in the value of the stationary 
equilibrium angle in small molecules. For significantly larger clusters, however, the effect 
of hydrodynamic interactions could be more complicated, and warrants further investigations. 
Allowing for particles to move in three dimensions does not change the global picture 
presented here. Geometric modifications will alter the thresholds for the transitions 
between isomers and new isomer configurations (e.g., a pyramid for $AB_3$) will emerge. 
However, we expect no new phenomenology to emerge due to the possibility of bending in 
the third direction, except for the formation of active chiral molecules. We relegate 
a detailed examination of the above effects to a future publication.

\acknowledgments
This research is supported by Fondecyt Grant No. 1100100 and Anillo grant ACT 127 (R.S.),
and Human Frontier Science Program (HFSP) grant RGP0061/2013 (R.G.).


\begin{thebibliography}{rsc}
\bibitem{note} 
Although we have used the language of diffusiophoresis throughout the Letter, the scheme presented 
here is generic for all phoretic transport mechanisms, as discussed in Ref. \cite{gla2007}.
\end{thebibliography}
\end{document}